%% file: bare_jrnl_new_sample4.tex
\documentclass[letterpaper, 10pt, conference]{ieeeconf}  
\IEEEoverridecommandlockouts                              

\usepackage[utf8]{inputenc}
\usepackage[T1]{fontenc}
\usepackage{textcomp}

\usepackage{amsmath,amsfonts}
\usepackage{algorithmic}
\usepackage{algorithm}
\usepackage{array}
\usepackage[caption=false,font=small]{subfig}
\usepackage{textcomp}
\usepackage{url}
\usepackage{verbatim}
\usepackage{graphicx}
\usepackage{cite}
\usepackage{color}
\usepackage[normalem]{ulem}
\usepackage{soul}
\usepackage{tikz}
\usepackage{latexsym}
\usepackage{amssymb}
\usepackage{mathtools}
\usepackage{cleveref}
\usepackage{xfrac}
\usepackage{lipsum}
\usepackage{float}
\usepackage{tabularx}
\usepackage[only,llbracket,rrbracket]{stmaryrd}
\usetikzlibrary {spy}
\makeatletter
\newcommand*{\centerfloat}{%
  \parindent \z@
  \leftskip \z@ \@plus 1fil \@minus \textwidth
  \rightskip\leftskip
  \parfillskip \z@skip}
\makeatother

\DeclareUnicodeCharacter{2212}{$-$}

\usepackage{tikz} 
\newcommand\copyrighttext{%
	\footnotesize \copyright 2025 IEEE. Personal use of this material is permitted. Permission from IEEE must be obtained for all other uses, in any current or future media, including reprinting/republishing this material for advertising or promotional purposes, creating new collective works, for resale or redistribution to servers or lists, or reuse of any copyrighted component of this work in other works.}
\newcommand\copyrightnotice{%
	\begin{tikzpicture}[remember picture,overlay]
		\node[anchor=south,yshift=10pt] at (current page.south) {\fbox{\parbox{\dimexpr\textwidth-\fboxsep-\fboxrule\relax}{\copyrighttext}}};
	\end{tikzpicture}%
}

\begin{document}

\title{Symbolic Control for Autonomous Docking of Marine Surface Vessels}

\author{Elizabeth Dietrich$^*$, Emir Cem Gezer$^*$, Bingzhuo Zhong$^{\dag}$, Murat Arcak, Majid Zamani, \\Roger Skjetne, and Asgeir Johan Sørensen
        \thanks{$^*$Both authors contributed equally. 
        $^{\dag}$ Corresponding Author.}
        \thanks{E. Dietrich and M. Arcak are with the 
        University of California, Berkeley, USA. Email: {\tt\small \{eadietri, arcak\}@berkeley.edu}.
        E. C. Gezer, R. Skjetne and A. J. Sørensen are with the 
        Norwegian University of Science and Technology, Trondheim, Norway. Email: {\tt\small \{emir.cem.gezer,roger.skjetne, asgeir.sorensen\}@ntnu.no}.
	B. Zhong is with the 
    Hong Kong University of Science and Technology (Guangzhou), China. Email: {\tt\small bingzhuoz@hkust-gz.edu.cn}.
    M. Zamani is with the 
    University of Colorado Boulder, USA. Email: {\tt\small majid.zamani@colorado.edu}.}
}

\markboth{}
{FirstAuthorSurname \MakeLowercase{\textit{et al.}}: ShortTitle} 

\maketitle

\begin{abstract}

We develop a hierarchical control architecture for autonomous docking maneuvers of a dynamic positioning vessel and provide formal safety guarantees. At the upper-level, we treat the vessel's desired surge, sway, and yaw velocities as control inputs and synthesize a symbolic controller in real-time. The desired velocities are then 
executed by the vessel's low-level velocity feedback control loop. 
We next investigate methods to optimize the performance of the proposed control scheme.
The results are evaluated on a simulation model of a marine surface vessel in the presence of static obstacles and, for the first time, through physical experiments on a scale model vessel.

\end{abstract}
\copyrightnotice

\vspace*{-4mm}
\section{Introduction}

Autonomy is critical 
for enhancing the efficiency and operational safety of marine surface vessels (MSV)~\cite{wrobel2017towards}.
Low-level automatic control functionalities, such as dynamic positioning~\cite{sorensen2011survey}, are well studied, but significant problems remain in providing formal guarantees for high-level control tasks like automatic path planning, collision avoidance, and guidance. 
Traditionally, these tasks have been performed by operators who rely on training, experience, situational awareness, and risk understanding. However, in 2023, 
the European Maritime Safety Agency, reported a total of 418 vessel collision occurrences, 
where human actions were identified as the primary cause, accounting for 58.4\% of the total \cite{emsa2024}. In contrast, autonomy leverages advanced sensing, computation, and control capabilities, offering enhanced safety, improved usage of human resources, reduced environmental impact, and minimized human injuries and fatalities.

\begin{figure}[!t]
	\centering
    \resizebox{.9\linewidth}{!}{
        \input{diagrams/overview} 
    }
	\caption{Symbolic control architecture for Dynamic Positioning (DP) vessels.
    }
	\label{fig:case1}
    \vspace{-.5cm}
\end{figure}

Docking is among the complex high-level control tasks for MSVs, which largely remains manually performed due to the high risk of collision and strict precision requirements~\cite{simonlexau2023automated}. When a vessel approaches a harbor for docking, it first enters the harbor basin where it must safely maneuver to a docking location while avoiding structures and other vessels. Existing autonomous docking methods use various  planning algorithms, such as Dubins curves~\cite{cai2023long}, 
A$^{*}$\cite{yuan2023event}, state-lattice\cite{bergman2020optimization}, and Voronoi diagrams \cite{schoener22}, combined with navigation systems, electronic nautical charts, and low-level control methods, including 
PID~\cite{8733620}, adaptive control~\cite{baek2022model}, linear quadratic regulator \cite{Tran2014ASB}, model predictive control~\cite{kockum2022autonomous}, and artificial neural networks~\cite{10.1007/978-3-031-08223-8_38}. However, due to the heuristic nature of many of these approaches, 
the safety of docking 
\emph{cannot} be verified a priori.

To address the unique safety challenges, we propose a \emph{correct-by-construction} control scheme using rigorous tools from control theory and formal methods~\cite{baier2008principles}. These tools, originating in computer science, are intended to ensure the correct operation of computer programs and digital circuits. Further, they offer formal, or mathematical, guarantees of correct behavior with respect to a formal system specification a priori---a significant improvement over the common practice of verification after design completion~\cite{YIN2024100940}, which typically necessitates extensive testing and redesign cycles~\cite{DNV2018}.

Among correct-by-construction synthesis methods, \emph{symbolic control} 
``abstracts'' the original continuous system into 
a \emph{symbolic system} with finite states and input sets. 
Controllers are then synthesized over the symbolic system to meet complex logical high-level control objectives and subsequently refined back to the original system.
Since the finite abstraction covers
all behaviors of the underlying dynamical system, this approach gives formal safety guarantees by construction.
Symbolic control has been
explored for 
linear systems~\cite{tabuada2009verification}, nonlinear systems with bounded disturbances~\cite{reissig2016feedback}, and nonlinear stochastic games~\cite{zhong2023automata}.

In this paper, we propose a hierarchical, {real-time} symbolic control scheme to address the safe autonomous docking problem for Dynamic Positioning (DP) vessels and present real-world experiments. At the upper-level, we leverage the vessel's kinematic model with desired surge, sway, and yaw velocities as control inputs, and synthesize a symbolic controller in real-time using the approach in~\cite{rungger2016scots}. At the lower-level, the desired velocities are executed by a low-speed velocity feedback control loop in the DP control system,  assigning control signals to the actuators of the vessel. 
As the vessel moves to the docking pose, it continuously updates obstacle positions 
and re-sythesizes the symbolic controller,
enabling the symbolic controller to make informed decisions.

A traditional shortcoming of symbolic control is that the finite state and input sets are created by partitioning the original sets with grids, adding to the computational complexity of this method and hampering practical applications.
To overcome this limitation and to enable real-time synthesis, 
we employ parallel computation with graphics processing units (GPUs) on a server running 
pFaces~\cite{khaled2019pfaces}.
We then identify the optimal control input from the synthesized action list.
The effectiveness of the proposed scheme is demonstrated on a simulation model of a MSV and --- for the first time --- through physical experiments on a scale model vessel. 
Additionally, we show that we can synthesize and re-synthesize a symbolic controller in real time, which is critical  for situational awareness in a dynamic environment. Experiments in a scaled environment are an important step in tuning and deploying software on real-world vessels \cite{gezer2025digital}; it allows us to close the sim-to-real gap efficiently and safely.

In earlier work~\cite{meyer2020continuous}, we investigated correct-by-construction synthesis for safe autonomous harbor maneuvering, but the  method was not amenable to real-time synthesis and
no physical experiments were conducted. The contributions of the present paper are: 1) to leverage
parallel computation and GPUs as proposed in~\cite{zhong2023towards,mahmoud2021efficient} to make real-time synthesis possible, allowing us to re-synthesize a symbolic controller in the presence of a dynamic environment, and 2) to demonstrate the results experimentally --- a crucial step towards deployment in real-world autonomous docking. The paper marks a new stage of maturity for symbolic control, demonstrating real-time synthesis capability, on a practical problem of great interest to the maritime industry.

\section{System Dynamics}\label{section2}

We use a three degrees of freedom (3-DOF) maneuvering ship model that considers surge, sway, and yaw motion~\cite{fossen2011handbook}. 
The kinematic model describes the relationship between the vessel’s body-fixed velocities and its position and orientation in the world-fixed frame. This relationship is expressed as
\begin{gather}
     \dot{\eta} = \mathbf{R}(\eta_\psi)\nu, \quad \mathbf{R}(\eta_\psi) = 
    \begin{bmatrix}
            \cos \eta_\psi & -\sin \eta_\psi & 0 \\
            \sin \eta_\psi & \cos \eta_\psi & 0 \\
            0 & 0 & 1
    \end{bmatrix},\label{eq:2_kinematic_model}
\end{gather}
where $\eta = [\eta_x, \eta_y, \eta_\psi]^\top$ represents the vessel's position $(\eta_x, \eta_y)$ and orientation $(\eta_\psi)$ in the world-fixed frame. Furthermore, $\nu = [\nu_x, \nu_y, \nu_\psi]^\top$ denotes the vessel’s linear and angular velocities in the body-fixed frame, corresponding to surge, sway, and yaw motions, respectively. $\mathbf{R}(\eta_\psi)$ is the rotation matrix that transforms velocities from the body-fixed frame to the world-fixed frame. In this work, linear velocities are expressed in meters per second ($m/s$) and angular velocities in radians per second ($rad/s$).

The kinetic model describes the forces and moments acting on the vessel and their effect on its velocity. The kinetics of the 3-DOF motion are given by
\begin{align}
    \mathbf{M}\dot{\nu} + \mathbf{C}(\nu)\nu + \mathbf{D}(\nu)\nu = \tau + b\label{eq:2_kinetic_model},
\end{align}
where $\mathbf{M} \in \mathbb{R}^{3\times3}$ is the inertia matrix, including the vessel’s mass and added mass terms, $\mathbf{C}(\nu) \in \mathbb{R}^{3\times3}$ represents Coriolis and centripetal forces, $\mathbf{D}(\nu) \in \mathbb{R}^{3\times3}$ is the hydrodynamic damping matrix accounting for drag forces, $\tau \in \mathbb{R}^3$ represents the control forces and moments, and $b \in \mathbb{R}^3$ represents slowly varying environmental or vessel loads.

\section{Control Strategy} \label{controlstrat}

We propose a hierarchical control architecture for a surface vessel that synthesizes a symbolic controller in real-time to enhance performance in autonomous 
docking maneuvers.
Fig. \ref{fig:case1} highlights the upper- and lower-level controllers of this architecture. 
Section \ref{symcontrol} describes the design of the symbolic controller that is deployed as the upper-level controller; Section \ref{DP} elaborates on the velocity feedback controllers applied in the low-level velocity control loop.

\subsection{Symbolic Control}\label{symcontrol}

Symbolic control \cite{10.5555/1717907} 
is an approach to derive provably safe controllers that satisfy complex logical specifications, including objectives and constraints in linear temporal logic. 
Symbolic control relies on the existence of a symbolic system, which is a finite abstraction of the continuous system. 
We use the approach in \cite{reissig2016feedback} to obtain a feedback refinement relation for the vessel, as modeled by (\ref{eq:2_kinematic_model}), that enables abstraction-based controller synthesis and leads to the control loop seen in Fig. \ref{fig:refine}. 

To better understand the correctness guarantees we achieve through this process, let $S_1$ and $S_2$ represent the plant (\ref{eq:2_kinematic_model}) and an abstraction of $S_1$, respectively, such that they are related through a feedback refinement relation, $Q$. Further, let $\Sigma_1$ be a specification on the state and input space and assume that $\Sigma_2$ is the corresponding abstract specification.
Then, $S_2$ 
contains the dynamical behaviors of $S_1$, and it is a
finite-state 
model that is amenable to
synthesis using standard search algorithms. 

Given this $Q$, the following holds: if the abstract controller $C$ solves the abstract control problem $(S_2, \Sigma_2)$, then the refined controller $C \circ Q$ solves the actual control problem $(S_1, \Sigma_1)$ [Theorem III.5, \cite{reissig2016feedback}]. Therefore, the resulting \textit{symbolic controller}, $C \circ Q$, is guaranteed to output a list of all possible control actions that satisfy the given specification. 
We optimize over this list to determine the optimal control input,
resulting in a velocity command executed by a low-level velocity controller, as shown in Fig. \ref{fig:case1}. 

\begin{figure}[ht]
    \centerfloat
    \resizebox{0.85\linewidth}{!}{
        \input{diagrams/symbolic_controller}
    }
    \caption{Closed loop: abstraction, refinement, and optimization. 
    }
    \label{fig:refine}
\end{figure}
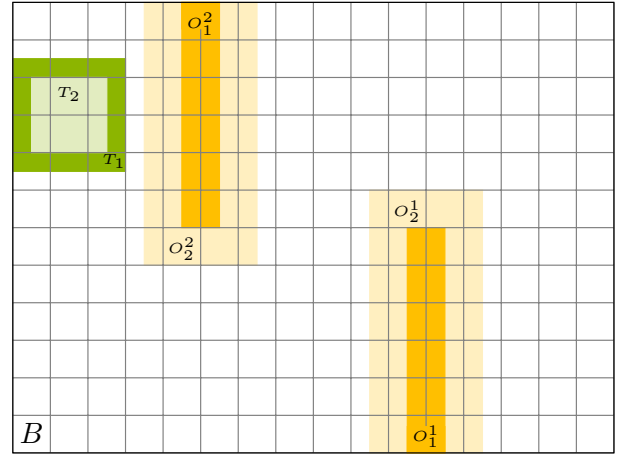
\begin{figure}[ht]
    \centerfloat
    \resizebox{0.8\linewidth}{!}{
        \input{diagrams/symbolic_system}
    }
    \caption{Discretized system with boundary $B =8 m \times 6 m$ and states of $S_1$ and $S_2$. $S_1$ has obstacles $O^1_{1}$ and $O^2_{1}$ (dark orange) and target docking position, $T_1$ (dark green). $S_2$ has obstacles $O^1_{2}$ and $O^2_{2}$ (light orange) and target docking position, $T_2$ (light green).}
    \label{fig:discrete_system}
\end{figure}

We adopt the convention where the states of system $i$ are denoted with subscript $i$ and $[a, b]$ denotes a discrete interval. We aim to guide the vessel (\ref{eq:2_kinematic_model}) throughout a region with boundary, $B$, into a target docking position, $T_1$, while avoiding obstacles $O^1_1$ and $O^2_1$ as seen in Fig. \ref{fig:discrete_system}. We dynamically measure the state of the vessel and obstacles. Then we use this information to perform the controller synthesis task that results in new control inputs that are sent to the vessel every $2$ seconds. This update frequency accounts for the settling time of the low-level velocity controller. We formulate our high-level control objective $(S_1, \Sigma_1)$ where $S_1$ is a mathematical representation of the physical vessel model and $\Sigma_1$ is the following specification:
\begin{align}
\label{sigma1}
        \{ (u, x) &\in (U_1 \times X_1)^{\mathbb{Z}_+} | x(0) \in B  \setminus (O^1_1, O^2_1)\\ 
    & \Rightarrow \forall_{t \in \mathbb{Z_+}} (x(t) \notin \{O^1_{1}, O^2_{1}\} \land \exists_{t^{'} \in [t; \infty)}x(t^{'}) \in T_{1})\}, \notag
\end{align} 
with $U_1$ as the set of inputs and $X_1$ as the set of vessel states.

$S_2$ is connected to the plant via a static quantizer, irrespective of the specification imposed on the plant. In defining $S_2$, we introduce margins for over-approximations of obstacles and an under-approximation for the target docking position. This can be seen in Fig. \ref{fig:discrete_system} as $O^1_{2}$, $O^2_{2}$, and $T_2$. These margins account for the geometrical shape of the vessel. Alternatively, they can be set according to other factors, e.g. ~\cite{dawson2020provably,axelrod2018provably}. Therefore, the abstract specification, $\Sigma_2$, of our system is (\ref{sigma1}) where we substitute $O^1_{1}, O^2_{1}, T_1$ with $O^1_{2}, O^2_{2}, T_2$, respectively.

\begin{figure}[tb]
    \centering
    \input{diagrams/transition}
    \caption{Possible transitions from vessel state (solid circle) to locations in the state space (dashed circles). The controller synthesis procedure will eliminate $u_2$ as a viable control action since it results in an unsafe maneuver (the vessel would collide with the orange obstacle). The list of possible control inputs returned would consist of $u_1, u_3, u_4, u_5,$ and $u_6$.}
    \label{fig:transition}
\end{figure}
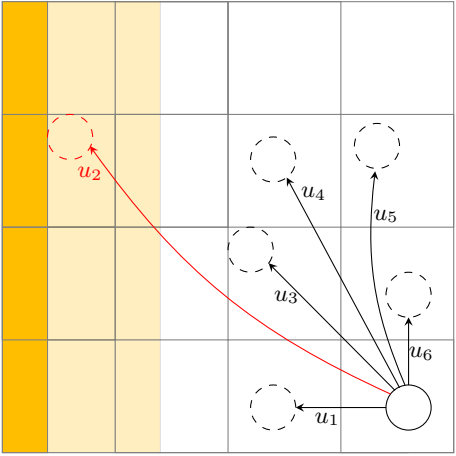

The abstract system, $S_2$, has an associated transition function 
among its states,
as illustrated in Fig. \ref{fig:transition}.
A standard search algorithm \cite{10.5555/1717907} is performed to determine safe control commands that allow the vessel to achieve its goal position. A list of all possible control actions that solve this problem are returned each time the synthesis task is performed with updated vessel and obstacle positions. 

Once a list of safe possible actions is computed, the plant may be given any one of the control commands. Every action on this list will result in safe progress towards the plant's goal. We optimize this selection process for realistic vessel behavior,
using a cost function that favors forward motion and reduces the norm between sequential control actions to reduce fluctuations in the control reference signal. Let $\mathcal{A} = \{\sigma_1, \sigma_2, \dots, \sigma_n\}$ be a list of actions where $\sigma_i \in \mathbb{R}^3$, $\sigma_{pre}$ be the previously chosen action, and $\Delta \sigma_i := \sigma_i - \sigma_{pre}$. We define the cost function $J(\sigma_i)$ as
\begin{align}
    J(\sigma_i) = \beta(\sigma_i)^\top W_i \beta(\sigma_i), \quad \beta(\sigma_i) = \begin{bmatrix}\sigma_i \\ \Delta \sigma_i\end{bmatrix} \in \mathbb{R}^6,
\end{align}
where $W \in \mathbb{R}^{6\times6}$ is positive semi-definite, as designed in Section \ref{optimization}; this matrix is application specific.
At every time step, we select the control action that minimizes the cost, $ \nu_{ref} := \sigma^* = \text{arg}\min_{\sigma_i\in \mathcal{A}}J(\sigma_i)$.

\subsection{Velocity Feedback Controller}\label{DP}

\begin{figure}[ht]
    \centering
    \resizebox{0.9\linewidth}{!}{
        \input{diagrams/dp_diagram}
    }
    \caption{The low-level velocity control loop of the DP control system.}
    \label{fig:dp_diagram}
\end{figure}
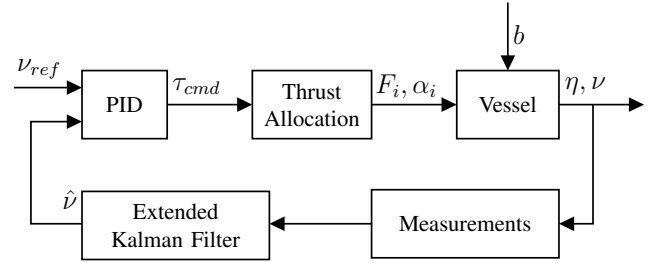

The velocity feedback controller, illustrated in Fig. \ref{fig:dp_diagram}, controls the vessel’s surge, sway, and yaw velocities, represented by $\nu_{ref}$, using a Multiple-Input Multiple-Output (MIMO) PID velocity controller, an extended Kalman filter (EKF) for state estimation, and a maneuvering based dynamic thrust allocation algorithm \cite{gezer2024maneuvering}. The PID controller calculates the desired force, $\tau_{cmd}$, to achieve the reference velocity $\nu_{ref}$. Then, the thrust allocation algorithm determines the necessary thruster forces, $F_1, F_2, F_3$, and azimuth angles, $\alpha_1, \alpha_2$, to apply these commands. By performing reference tracking, the PID controller smooths the discretized reference signal sent by the symbolic controller. Position and velocity predictions are computed using the \texttt{robot\_localization} package's EKF implementation, as described in \cite{MooreStouchKeneralizedEkf2014}. The measurement sources vary based on setup, as detailed in Section \ref{experimental_setup}.

\section{Experimental Setup}\label{experimental_setup}

\subsection{Problem Formulation}
We refine the definition of our symbolic control problem \eqref{sigma1} based on preliminary experimental observations. With the goal of ensuring reliable performance, we tracked the response time of our low-level velocity controller and defined the set of possible control inputs to be the best-performing combinations of linear and angular velocities while accounting for the testing arena 
and vessel capabilities. This resulted in the set of inputs, $U_1$, and vessel states, $X_1$,  defined as,
\begin{align}
    \begin{aligned}
    U_1 & = [-0.1, 0.2] \times [-0.1, 0.1] \times [-0.2, 0.2] \\
    X_1 & = [-3.5, 4.5] \times [-3.0, 3.0] \times [-\pi, \pi).
    \end{aligned}
\end{align}

\subsection{Simulation}
Preliminary testing was conducted in a 3-DOF simulation environment developed for a similar-sized vessel, Cybership (C/S) Enterprise I\cite{skaatun2011development}.
C/S Enterprise I is a 1:50 scale tugboat, with dimensions and weight comparable to those of C/S Voyager.
The simulation is based on the kinematic and kinetic equations \eqref{eq:2_kinematic_model} and \eqref{eq:2_kinetic_model}, and uses the Forward Euler method for numerical integration over time. The update equations for the vessel's pose and velocity are given as
\begin{subequations} \label{eq:simulation_integrator}
    \begin{align}
        \eta_{k+1} &= \eta_{k} + \Delta t \cdot \mathbf{R}(\eta_{k,\psi}) \nu_k, \\
        \nu_{k+1} &= \nu_{k} + \Delta t \cdot \mathbf{M}^{-1}(\tau_k - \mathbf{C}(\nu_k)\nu_k - \mathbf{D}(\nu_k)\nu_k),
    \end{align}
\end{subequations}
where $\Delta t$ represents the time step. Hydrodynamic coefficients for $\mathbf{M}$, $\mathbf{C}(\nu)$, and $\mathbf{D}(\nu)$ were obtained via system identification in \cite{skaatun2011development}. The simulation environment is implemented using Robot Operating System 2 (ROS 2) and runs in real-time. It models the vessel's surge, sway, and yaw motions by integrating the forces and torques applied by multiple thrusters, utilizing the integrator in \eqref{eq:simulation_integrator}. The simulator updates the vessel's position ($\eta$) and velocity ($\nu$) based on thrust commands received from ROS topics and publishes motion data in common ROS interfaces. 

\subsection{Towing Tank Environment}

Experimental trials were conducted in NTNU's Marine Cybernetics Laboratory (MCLab) using a scaled model vessel, the C/S Voyager. The towing tank has dimensions of $40 m \times 6.45 m \times 1.5 m$, and is equipped with a real-time positioning system, Qualisys, which includes Oqus cameras and the Qualisys Track Manager software. This environment is ideal for testing motion control systems for marine vessels, due to its manageable size and advanced instrumentation.

\begin{figure}[tbp]
    \centering
    \subfloat[\label{fig:arena_picture} Marine Cybernetics Laboratory
    ]{
        \includegraphics[width=0.7\linewidth, height=0.6\textheight, keepaspectratio]{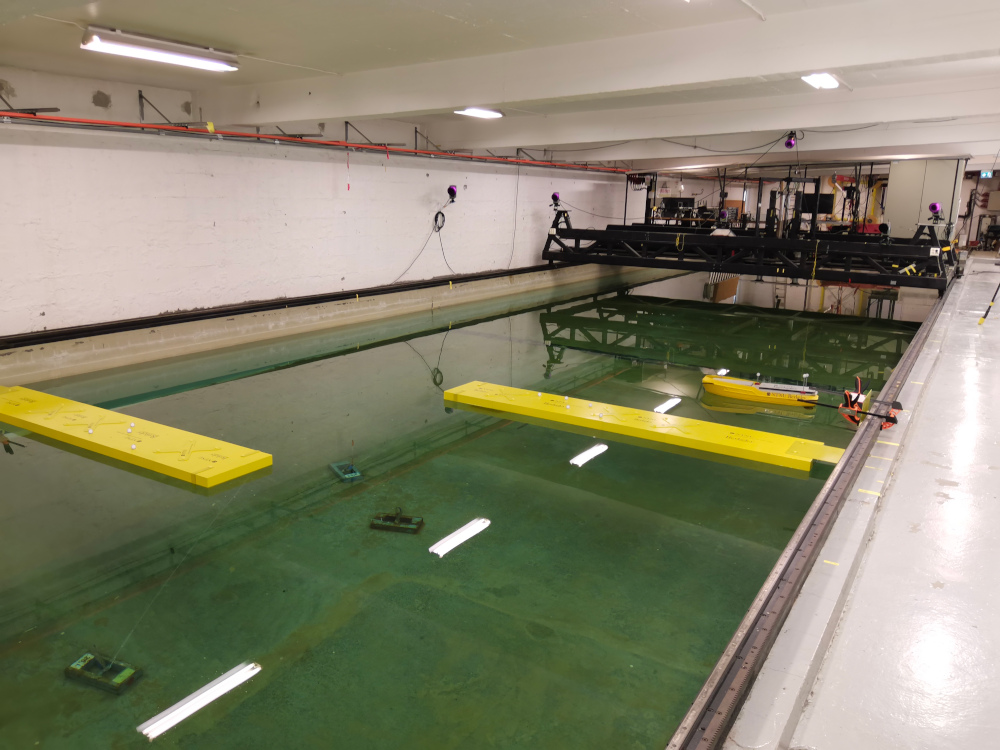}

    }
    \\
    \subfloat[\label{fig:voyager} C/S Voyager
    ]{
       \includegraphics[width=0.7\linewidth, height=0.6\textheight, keepaspectratio]{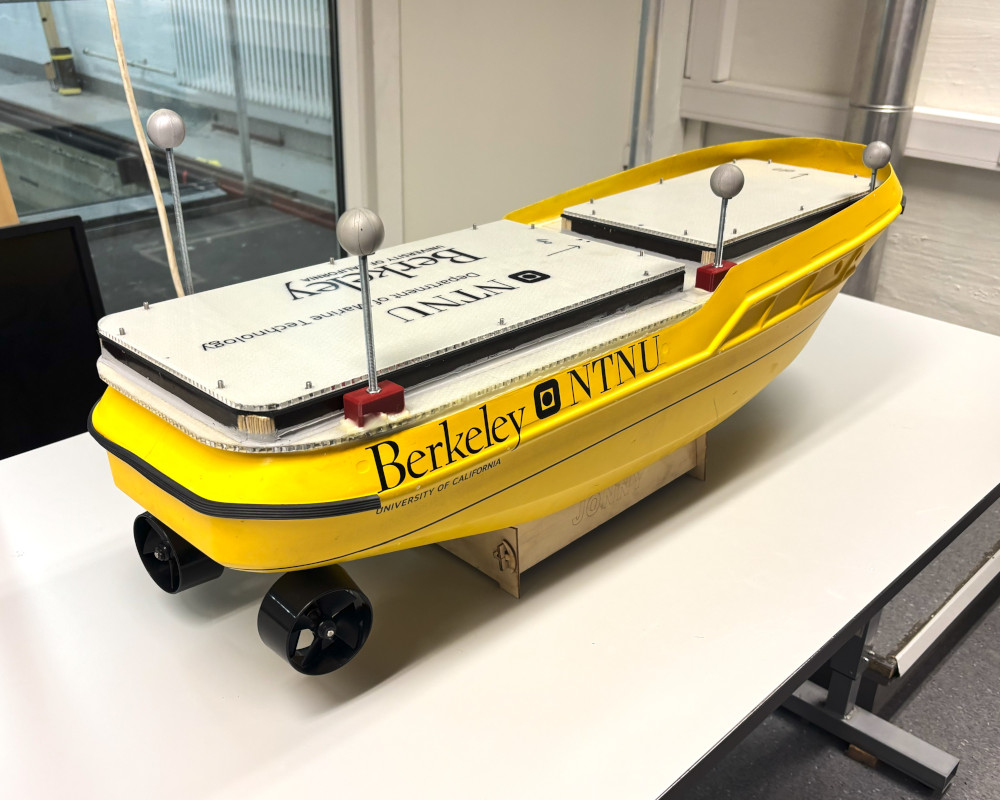}

    }
    \caption{ \protect\subref{fig:arena_picture} 
    Marine Cybernetics Laboratory (MCLab) where experiments took place in a towing tank ($40 m \times 6.45 m \times 1.5 m$) with two yellow, stationary obstacles ($0.5 m \times 3 m$) attached to the walls and a scaled model vessel (length $1 m$).\protect\subref{fig:voyager} Model vessel that was used in the experiments}
    \label{fig:setup}
\end{figure}

The main test involved two stationary obstacles for the vessel to evade while approaching the target docking position. The testing region for this experiment was a $8m \times 6m \times 1.5m$ portion of the towing tank. In this region, we used 7 motion capturing cameras, ensuring full coverage of the testing area. The two stationary obstacles, each $0.5 m \times 3 m$, were attached to the walls of the towing tank as depicted in Fig. \ref{fig:arena_picture}. No external forces (e.g. waves) were introduced to the tank during testing.

\subsection{Model Vessel}

The C/S Voyager, shown in Fig. \ref{fig:voyager}, is a 1:32 scale tugboat built at NTNU's MCLab to test advanced navigation and control systems. 
The C/S Voyager has over 4 hours of endurance and azimuth thrusters, which can reach a maximum force of more than 80N.
The vessel's components achieve real-time communication through ROS 2, which is installed on the onboard computer, a Raspberry Pi 4. The C/S Voyager has an array of sensing devices, including infrared spheres for accurate position tracking with the Qualisys motion capture system and Inertial Measurement Units (IMUs) for measuring linear accelerations and angular velocities. 

\subsection{Optimization Tuning}
\label{optimization}
Since the controller synthesis utilizes a simplified vessel dynamics model, some of the synthesized actions may not reflect realistic vessel behavior. For example, it may be safe for the vessel to perform docking backwards; however, it is impractical to exhibit this behavior in the physical world. The matrix $W$, introduced in Section \ref{symcontrol}, ensures that only reasonable actions are chosen. We designed $W$ to ensure the optimal control action was chosen from the synthesized action set. We derive $W$ systematically using domain specific knowledge and design iterations. $W$ is chosen conditionally based on the forward velocity component of the proposed control action, $\sigma_{i_{ \nu_x}}$, such that
\begin{align}
    W_i = \begin{cases}
        \text{diag}(7.5, 3, 1, 1, 1, 2.5)\quad  &\sigma_{i_{ \nu_x}} < 0\\
        \text{diag}(1,6,1,1,1,2.5)\quad &\text{otherwise.}
    \end{cases}
    \label{eq:experiment_weights}
\end{align}
Given a proposed action with a negative forward velocity, we penalize this backward motion (weight 7.5), discourage sway motion (weight 3), and minimize drastic changes in the yaw (weight 2.5). This ensures that the vessel prioritizes forward motion and limits usage of the tunnel thruster. Additionally, \eqref{eq:experiment_weights} minimizes rapid rotations of the vessel, reducing drastic changes in the heading. 
When a proposed action has a non-negative forward velocity, we only penalize sway motion and changes in the yaw. However, we must increase the sway penalty (weight 6), as the synthesized controller is more likely to suggest sway commands when we  make forward progress. We similarly discourage vessel rotations and changes in the yaw (weight 2.5).
\section{Results}\label{results}

\begin{figure}[htp]
    \centering
    \subfloat[\label{fig:sim_run} 
    ]{
        \includegraphics[width=0.8\linewidth, height=0.8\textheight, keepaspectratio]{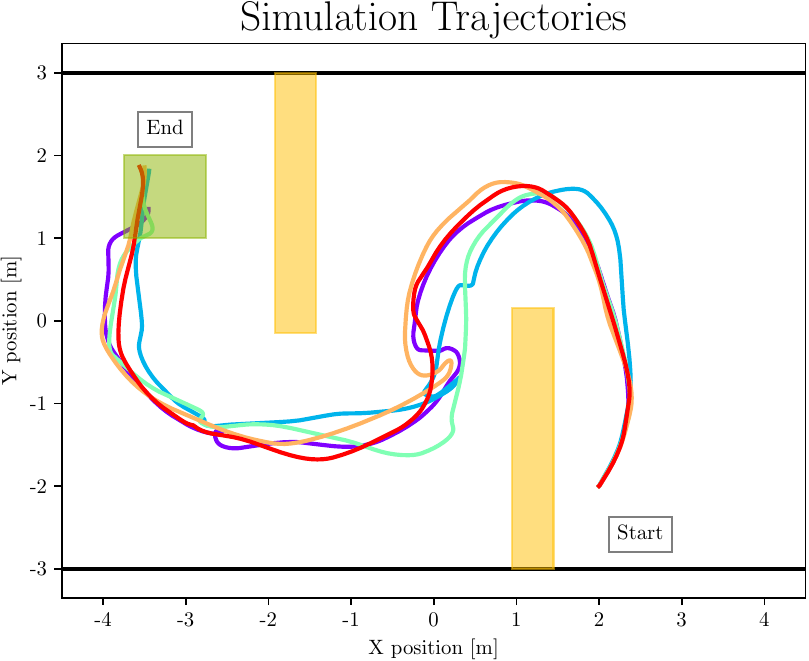}

    }
    \\
    \subfloat[\label{fig:successful_runs} 
    ]{
        \includegraphics[width=0.8\linewidth, height=0.8\textheight, keepaspectratio]{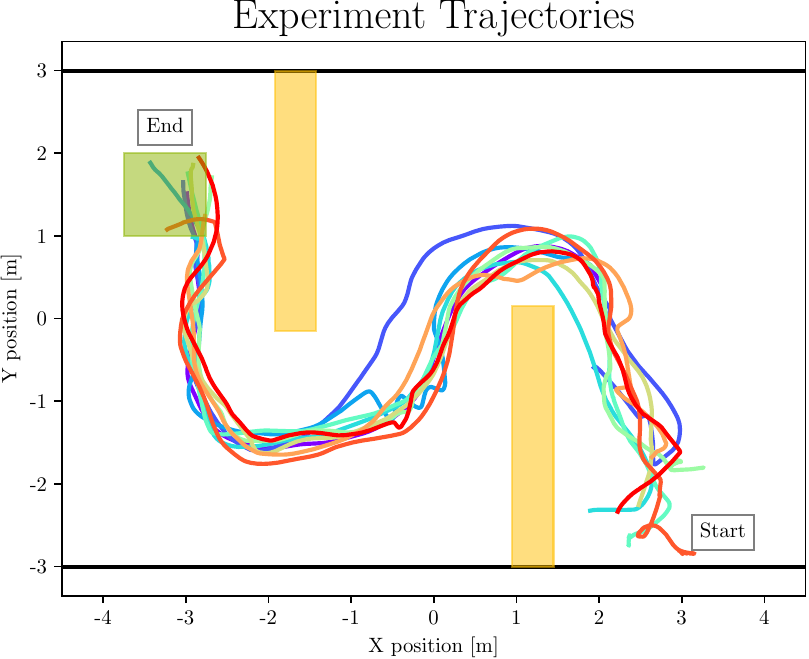}

    }
    \caption{ The yellow rectangles portray 3 meter obstacles, and the green box denotes the 1 meter square target. \protect\subref{fig:sim_run} 
    Trajectories of 5 successful docking maneuvers in the simulation environment, all starting from the same vessel position.
    \protect\subref{fig:successful_runs} Trajectories of 10 successful autonomous docking maneuvers in the real-world, starting from a range of vessel positions.}
    \label{fig:setup}
\end{figure}

To test the efficacy of our control strategy on both one and two obstacles before deploying to the real vessel, we performed experimental trials in simulation, mirroring the setup shown in Fig. \ref{fig:arena_picture}. We were able to successfully synthesize control strategies that safely traversed static obstacles in simulation, as seen in Fig. \ref{fig:sim_run} \footnote{Docking  in simulation environment: \url{https://youtu.be/TFynNTElZV8}}.

To execute symbolic controller synthesis in real-time, we utilized the pFaces \cite{khaled2019pfaces} accelerator deployed on a server equipped with an Nvidia A6000 GPU (48 GB of VRAM), 256 GB of memory, and an Intel Xeon 6226R processor with the pFaces Symbolic Control kernel~\cite{EP3633468A1}. The server performs the correct-by-construction synthesis task, as detailed in Section \ref{symcontrol}, in approximately 0.5 seconds, enabling us to perform controller re-synthesis in real-time.
To enable real-time interaction with the autonomous vessel, we developed a ROS 2 Python package that served as an interface to communicate control commands and vessel state. This architecture, as outlined in Fig. \ref{fig:case1}, provides a robust and scalable computational infrastructure for the iterative development, validation, and real-time implementation of advanced control strategies.

State information was relayed to the server after every completed synthesis task. The server allocated apx. 400MB of VRAM and apx. 500MB of system memory to complete the synthesis task. The necessary subsystems, such as the observers and DP controllers described in Fig. \ref{fig:dp_diagram}, were run on the onboard single board computer, a Raspberry Pi 4 with 8GB memory. The symbolic controller assumes: (1) the low-level control system is able to realize any commanded velocity and (2) the kinematic model, which the symbolic control is based upon, encompasses all possible behaviors of the real-world vessel. We observed this to be true in simulation, Eq. (\ref{eq:simulation_integrator}), and during the real-world experiments. Given these assumptions, the synthesized velocity commands are theoretically guaranteed to maintain vessel safety. We discuss discrepancies between these theoretical guarantees and the practical implementation in Section \ref{discussion}. 

The trajectories of the successful autonomous docking maneuvers, completed in the laboratory environment with the scale model vessel, are shown in Fig. \ref{fig:successful_runs}\footnote{ Docking of model vessel in towing tank: \url{https://youtu.be/q-qohEJciU4}}.
The vessel maneuvered generously around the obstacles, allowing extra space to maintain safety and ensure there would be no collisions.
On average, the vessel completed this trajectory in 2.67 minutes; the minium and maximum durations were 1.71 and 3.66 minutes, respectively.

\section{Discussion and Future Work}
\label{discussion}
 
We analyzed the low-level controller's ability to realize the synthesized velocity commands.
To quantify the error between commanded and realized velocities, we calculated the mean squared error (MSE), $\nu_{\text{\tiny{MSE}}} = [0.00768 \sfrac{m}{s}, 0.00431 \sfrac{m}{s}, 0.62108 \sfrac{deg}{s}]^\top$. 
Although we did not formally account for the tracking error,
it is possible to do so by performing a continuous abstraction in which the low-level controller guarantees a bound on the tracking error and the high-level symbolic synthesis accounts for the error, as proposed in~\cite{meyer2020continuous}. Replacing the existing low-level DP controller in experiments with one obtained from a continuous abstraction is left for future research.
Additionally, we employed a simplified vessel model, which neglects environmental loads, such as wind and waves, and we relied on vessel-server communication. In future work, we hope to address these limitations by developing the current vessel model through physics-informed system identification, introducing wave and wind models, and advancing the hardware capabilities of the C/S Voyager to enable experimental testing in outdoor environments. Further, we hope to investigate the extension of this approach to full-scale vessels.

\section{Conclusion}
We presented a hierarchical symbolic control strategy for  safe autonomous docking.
We utilized parallel computing and GPU acceleration to achieve real-time synthesis in dynamic settings, with experimental validation marking a key step toward real-world implementation. The work represents a new level of maturity in symbolic control, demonstrating real-time synthesis despite past computational limitations.

\section*{Acknowledgments}
This paper is supported in part by 
the NSF 
project CNS-2111688, NTNU VISTA Centre for Autonomous Robotic Operations Subsea (CAROS), 
Research Council of Norway (RCN) through SFI AutoShip (RCN project 309230), 
Guangzhou-HKUST(GZ) Joint Funding Program (Grant No.2025A03J4493), Education Bureau of Guangzhou Municipality,
and a grant from the Peder Sather Center for Advanced Study at UC Berkeley. The first author was also supported by a NSF Graduate Research Fellowship. 
The authors thank Emil Bratlie, Robert Opland, Terje Rosten, Vebjørn Steinsholt, Gabriel Bjørgan Vikan, and Anne Elise Havmo 
of the MCLab. 
\bibliography{sample-base} 
\bibliographystyle{IEEEtran}

\end{document}

%% file: diagrams/overview.tex
\begin{tikzpicture}[auto, scale=.65]
	\tikzset{
		mynode/.style={rectangle,rounded corners,draw=black,fill=TUMBlue!80,very thick, inner sep=1em, minimum size=3em, text centered, minimum width=10cm, text=white, minimum height=1.5cm},
		myarrow/.style={
			->,
			thick,
			shorten <=2pt,
			shorten >=2pt,},
		mylabel/.style={text width=7em, text centered} 
	} 
\definecolor{ao}{rgb}{0.0, 0.5, 0.0}
\definecolor{corn}{rgb}{0.98, 0.93, 0.36}
\definecolor{aliceblue}{rgb}{0.94, 0.97, 1.0}

\tikzstyle{every node}=[font=\normalsize]
\draw [draw=black, fill=corn, fill opacity=0.25, rounded corners=10] (4.25,14.75) rectangle (16.25,8.25);
\draw [draw=black] (7, 8.4) node [above, text=black]{\small{\textbf{Upper-Level Controller}}};

\draw [draw=black, fill=ao, fill opacity=0.15, rounded corners=10] (10,9.75) rectangle ++(6,4.75);
\draw [draw=black] (12.75, 13.65) node [above, text=black]{\small{\textbf{Lower-Level Controller}}};

\draw [fill=white, rounded corners = 7.2] (10.5,11.4) rectangle  ++(2.25,2.25);
\draw [draw=black] (11.62, 12.6) node [above, text=black]{\footnotesize{Slow-Speed}};
\draw [draw=black] (11.62, 12.15) node [above, text=black]{\footnotesize{Velocity}};
\draw [draw=black] (11.62, 11.8) node [above, text=black]{\footnotesize{Controller}};
\draw [draw=black] (12.75, 12.9) node [right, text=black]{\scriptsize{low-level control}};
\draw [draw=black] (12.75, 12.5) node [right, text=black]{\scriptsize{inputs}};

\draw [draw=black, -stealth] (12.75,13.15) -- (16.6,13.15);
\node[inner sep=0pt] () at (18, 13.15)
    {\includegraphics[width=.1\textwidth]{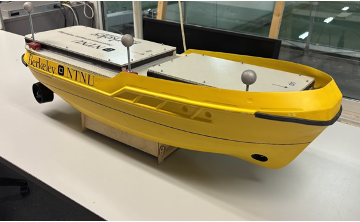}};

\draw [fill=white, rounded corners = 7.2] (17, 8.6) rectangle  ++(2,2); 
\draw [draw=black] (18, 9.5) node [above, text=black]{\footnotesize{State}};
\draw [draw=black] (18, 9.1) node [above, text=black]{\footnotesize{Observer}};

\draw [draw=black] (17,9.5) -- (5.7,9.5);
\draw [draw=black, -stealth] (5.7,9.5) -- (5.7,11.35);
\draw [draw=black] (5.8, 10.15) node [right, text=black]{\scriptsize{position, heading,}};
\draw [draw=black] (5.8, 9.75) node [right, text=black]{\scriptsize{and obstacles}};

\draw [draw=black, -stealth] (18.05,12.3) -- (18.05,10.6);

\draw [fill=white, rounded corners = 7.2] (4.6,11.35) rectangle  ++(2.25,2.25);
\draw [draw=black] (5.75, 12.3) node [above, text=black]{\footnotesize{Symbolic}};
\draw [draw=black] (5.75, 11.9) node [above, text=black]{\footnotesize{Controller}};

\draw [draw=black] (17,10) -- (11.6,10);
\draw [draw=black, -stealth] (11.6,10) -- (11.6,11.4);
\draw [draw=black] (11.7, 10.65) node [right, text=black]{\scriptsize{surge, sway and }};
\draw [draw=black] (11.7, 10.25) node [right, text=black]{\scriptsize{yaw velocities}};

\draw [draw=black, -stealth] (6.85,13.15) -- (10.5,13.15);
\draw [draw=black] (8.44, 13.15) node [below, text=black]{\scriptsize{desired surge, sway}};
\draw [draw=black] (8.44, 12.75) node [below, text=black]{\scriptsize{and yaw velocities}};

\draw [fill=aliceblue, rounded corners = 7.2] (4.25,15) rectangle ++(6,2);
\draw [draw=black, stealth-] (5.7,13.6) -- (5.7,15);
\draw [draw=black, stealth-] (10.25,16) -- (11,16);
\draw [draw=black] (6.5, 16.5) node [text=black]{\scriptsize{Real-time controller}};
\draw [draw=black] (6.5, 16.1) node [text=black]{\scriptsize{synthesis on a}};
\draw [draw=black] (6.5, 15.7) node [text=black]{\scriptsize{server with GPUs}};
\node[inner sep=0pt] () at (9.25, 16)
    {\includegraphics[width=.05\textwidth]{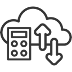}};

\draw [draw=black] (18.05,14) -- (18.05,16);
\draw [draw=black, -stealth] (18.05,16) -- (16,16);
\draw [ rounded corners = 7.2, dashed] (11,15) rectangle ++(5,2);
\draw [draw=black] (13.5, 16.4) node [text=black]{\scriptsize{Complex logical control}};
\draw [draw=black] (13.5, 16.0) node [text=black]{\scriptsize{objectives, including}};
\draw [draw=black] (13.5, 15.6) node [text=black]{\scriptsize{autonomous docking, etc.}};

\end{tikzpicture}

%% file: diagrams/symbolic_controller.tex
\tikzset{every picture/.style={line width=0.75pt}} 

\begin{tikzpicture}[x=0.75pt,y=0.75pt,yscale=-1,xscale=1]

\draw  [dash pattern={on 0.84pt off 2.51pt}] (235,135) -- (485,135) -- (485,240) -- (235,240) -- cycle ;
\draw   (245,160) -- (335,160) -- (335,230) -- (245,230) -- cycle ;
\draw   (350,145) -- (440,145) -- (440,185) -- (350,185) -- cycle ;
\draw  [dash pattern={on 0.84pt off 2.51pt}] (405,196.9) -- (485,196.9) -- (485,240) -- (405,240) -- cycle ;
\draw    (245,200) -- (215,200) -- (200,200) -- (185,200) -- (170,200) -- (170,183) ;
\draw [shift={(170,180)}, rotate = 90] [fill={rgb, 255:red, 0; green, 0; blue, 0 }  ][line width=0.08]  [draw opacity=0] (8.93,-4.29) -- (0,0) -- (8.93,4.29) -- cycle    ;
\draw    (350,105) -- (395,105) -- (395,120) -- (395,142) ;
\draw [shift={(395,145)}, rotate = 270] [fill={rgb, 255:red, 0; green, 0; blue, 0 }  ][line width=0.08]  [draw opacity=0] (8.93,-4.29) -- (0,0) -- (8.93,4.29) -- cycle    ;
\draw    (395,185) -- (395,200) -- (370,200) -- (338,200) ;
\draw [shift={(335,200)}, rotate = 360] [fill={rgb, 255:red, 0; green, 0; blue, 0 }  ][line width=0.08]  [draw opacity=0] (8.93,-4.29) -- (0,0) -- (8.93,4.29) -- cycle    ;
\draw   (260,70) -- (350,70) -- (350,125) -- (260,125) -- cycle ;
\draw   (125,126) -- (215,126) -- (215,180) -- (125,180) -- cycle ;
\draw    (170,125) -- (170,105) -- (185,105) -- (257,105) ;
\draw [shift={(260,105)}, rotate = 180] [fill={rgb, 255:red, 0; green, 0; blue, 0 }  ][line width=0.08]  [draw opacity=0] (8.93,-4.29) -- (0,0) -- (8.93,4.29) -- cycle    ;

\draw (290,195) node   [align=left] {\begin{minipage}[lt]{44.68pt}\setlength\topsep{0pt}
\begin{center}
abstract\\controller\\$\displaystyle S_{2}$
\end{center}

\end{minipage}};
\draw (305,97.5) node   [align=left] {\begin{minipage}[lt]{31.64pt}\setlength\topsep{0pt}
\begin{center}
vessel\\$\displaystyle S_{1}$
\end{center}

\end{minipage}};
\draw (397,164.5) node   [align=left] {\begin{minipage}[lt]{44.69pt}\setlength\topsep{0pt}
\begin{center}
quantizer
\end{center}

\end{minipage}};
\draw (445,218.45) node   [align=left] {\begin{minipage}[lt]{44.68pt}\setlength\topsep{0pt}
\begin{center}
symbolic\\controller
\end{center}

\end{minipage}};
\draw (210.5,94.5) node   [align=left] {\begin{minipage}[lt]{61.69pt}\setlength\topsep{0pt}
\begin{center}
velocity input
\end{center}

\end{minipage}};
\draw (393.5,94.5) node   [align=left] {\begin{minipage}[lt]{56.59pt}\setlength\topsep{0pt}
\begin{center}
vessel state
\end{center}

\end{minipage}};
\draw (170,153) node   [align=left] {\begin{minipage}[lt]{57.15pt}\setlength\topsep{0pt}
\begin{center}
optimization\\tuning
\end{center}

\end{minipage}};

\end{tikzpicture}

%% file: diagrams/symbolic_system.tex
\def\vessel#1#2#3#4#5#6{
	\begin{scope}[shift={#1}, rotate=#2, scale=#3]
		\draw [fill=#4,draw=#5, opacity=#6](-1,1) -- (0,2) -- (1,1) -- (1,-1) -- (-1,-1) -- cycle;
	\end{scope}
}

\definecolor{Amber}{rgb}{1.0, 0.75, 0.0}
\definecolor{applegreen}{rgb}{0.55, 0.71, 0.0}

\begin{tikzpicture}[node distance=4cm, auto, scale=.90]  
	\tikzset{
		mynode/.style={rectangle,rounded corners,draw=black,fill=TUMBlue!80,very thick, inner sep=1em, minimum size=3em, text centered, minimum width=10cm, text=white, minimum height=1.5cm},
		myarrow/.style={
			->,
			thick,
			shorten <=2pt,
			shorten >=2pt,},
		mylabel/.style={text width=7em, text centered} 
	}

	
        \draw[draw=applegreen, fill=applegreen] (0,3.75) rectangle ++(1.5, 1.5);
        \draw[draw=white, fill=white] (0.25,4) rectangle ++(1, 1);
        \draw[draw=applegreen, fill=applegreen, opacity=0.25] (0.25,4) rectangle ++(1, 1);
        
        \draw[draw=Amber, fill=Amber] (2.25,3) rectangle ++(0.5, 3);
        \draw[draw=Amber, fill=Amber, opacity=0.25] (1.75,2.5) rectangle ++ (1.5, 3.5);
        \draw[draw=Amber, fill=Amber] (5.25,0) rectangle ++(0.5, 3);
        \draw[draw=Amber, fill=Amber, opacity=0.25] (4.75,0) rectangle ++ (1.5, 3.5);
        \draw[gray, step=0.5] (0,0) grid (8,6);
        \draw[draw=Amber, fill=Amber] (5.25, 0) rectangle ++(0.5, 0.35);
        \draw[draw=black](5.5, .5) node [below]{\tiny{$O^1_1$}};
        \draw[draw=black](5.25, 3.5) node [below]{\tiny{$O^1_2$}};
        
        \draw[draw=Amber, fill=Amber] (2.25, 5.65) rectangle ++(0.5, 0.35);
        \draw[draw=black](2.5, 6) node [below]{\tiny{$O^2_1$}};
        \draw[draw=black](2.25, 3) node [below]{\tiny{$O^2_2$}};

        \draw[draw=black](0.75, 4.55) node [above]{\tiny{$T_2$}};
        \draw[draw=black](1.35, 3.65) node [above]{\tiny{$T_1$}};
        \draw[draw=black](0.25, 0) node [above]{$B$};
        \draw[draw=black] (0,0) rectangle ++(8,6);

\end{tikzpicture} 

%% file: diagrams/transition.tex
\usetikzlibrary{arrows.meta, bending, shapes.misc}

\begin{tikzpicture}[]
\tikzset{
		mynode/.style={
            rectangle,
            rounded corners,
            draw=black,
            fill=TUMBlue!80,
            very thick, 
            inner sep=1em, 
            minimum size=3em, 
            text centered, 
            minimum width=10cm, 
            text=white, 
            minimum height=1.5cm
        },
		myarrow/.style={
			->,
			thick,
			shorten <=2pt,
			shorten >=2pt,
        },
		mylabel/.style={
            text width=7em, 
            text centered
        } 
	} 
 
\draw[draw=gray, fill={rgb,255:red,255; green,255; blue,255}];
\draw[draw=gray]  (5,16.25) rectangle (10,11.25);
\draw[draw=gray]   (5,16.25) rectangle (7.5,13.75);
\draw[draw=gray]   (7.5,16.25) rectangle (8.75,15);
\draw[draw=gray]   (8.75,15) rectangle (10,13.75);
\draw[draw=gray]   (6.25,15) rectangle (7.5,13.75);
\draw[draw=gray]   (5,13.75) rectangle (6.25,12.5);
\draw[draw=gray]   (6.25,12.5) rectangle (7.5,11.25);
\draw[draw=gray]   (7.5,11.25) rectangle (9.5,11.25);
\draw[draw=gray]   (7.5,13.75) rectangle (8.75,12.5);
\draw[draw=gray]   (8.75,12.5) rectangle (10,11.25);
\draw[draw=gray]   (6.25,16.25) rectangle (7.5,15);
\draw [draw=gray, fill={rgb,255:red,255; green,191; blue,0} ] (5,16.25) rectangle (5.5,11.25);
\draw[draw=gray]  (5,15) rectangle (6.25,13.75);
\draw[draw=gray]  (5,13.75) rectangle (6.25,12.5);
\draw [draw=gray, fill={rgb,255:red,255; green,191; blue,0}, opacity=0.25 ] (5.5,16.25) rectangle (6.25,15);
\draw [draw=gray, fill={rgb,255:red,255; green,191; blue,0}, opacity=0.25 ] (6.25,16.25) rectangle (6.75,15);
\draw [draw=gray, fill={rgb,255:red,255; green,191; blue,0}, opacity=0.25 ] (5.5,15) rectangle (6.25,13.75);
\draw [draw=gray, fill={rgb,255:red,255; green,191; blue,0}, opacity=0.25 ] (6.25,15) rectangle (6.75,13.75);
\draw [draw=gray, fill={rgb,255:red,255; green,191; blue,0}, opacity=0.25 ] (5.5,13.75) rectangle (6.25,12.5);
\draw [draw=gray, fill={rgb,255:red,255; green,191; blue,0}, opacity=0.25 ] (6.25,13.75) rectangle (6.75,12.5);
\draw [draw=gray, fill={rgb,255:red,255; green,191; blue,0}, opacity=0.25 ] (5.5,12.5) rectangle (6.25,11.25);
\draw [draw=gray, fill={rgb,255:red,255; green,191; blue,0}, opacity=0.25 ] (6.25,12.5) rectangle (6.75,11.25);
\draw [draw=black] (9.5,11.75) circle (0.25cm);

\draw [draw=black, -stealth] (9.25,11.75) -- (8.25,11.75);
\draw [draw=black] (8.6, 11.8) node [below, text=black]{\small{$u_1$}};

\draw [draw=black, dashed] (8,11.75) circle (0.25cm);
\draw [draw=black, dashed] (9.5,13) circle (0.25cm);

\draw [draw=black, -stealth] (9.5,12) -- (9.5,12.75);
\draw [draw=black] (9.4, 12.35) node [right, text=black]{\small{$u_6$}};

\draw [draw=black, dashed] (7.75,13.5) circle (0.25cm);

\draw [draw=black, -stealth] (9.35,11.94) -- (7.95,13.35);
\draw [draw=black] (8.15, 13.15) node [below, text=black]{\small{$u_3$}};

\draw [draw=red, -stealth] (9.3,11.9) to [bend left=15] (5.97,14.65); 
\draw [draw=red] (5.97, 14.52) node [below, text=red]{\small{$u_2$}};

\draw [draw=red, dashed] (5.75,14.75) circle (0.25cm);
\draw [draw=black, dashed] (8,14.5) circle (0.25cm);
\draw [draw=black, dashed] (9.15,14.65) circle (0.25cm);

\draw [draw=black, -stealth] (9.4,11.97) -- (8.15,14.3);
\draw [draw=black] (8.45, 13.95) node [above, text=black]{\small{$u_4$}};

\draw [draw=black, -stealth] (9.46,12) to [bend left=15] (9.13,14.37);
\draw [draw=black] (9.25, 13.7) node [above, text=black]{\small{$u_5$}};
\end{tikzpicture}

%% file: diagrams/dp_diagram.tex
\tikzset{every picture/.style={line width=0.75pt}} 

\begin{tikzpicture}[x=0.75pt,y=0.75pt,yscale=-1,xscale=1]

\draw   (150,80) -- (200,80) -- (200,120) -- (150,120) -- cycle ;
\draw    (110,90) -- (147,90) ;
\draw [shift={(150,90)}, rotate = 180] [fill={rgb, 255:red, 0; green, 0; blue, 0 }  ][line width=0.08]  [draw opacity=0] (8.93,-4.29) -- (0,0) -- (8.93,4.29) -- cycle    ;
\draw   (250,80) -- (320,80) -- (320,120) -- (250,120) -- cycle ;
\draw   (370,70) -- (430,70) -- (430,130) -- (370,130) -- cycle ;
\draw   (150,151) -- (260,151) -- (260,191) -- (150,191) -- cycle ;
\draw    (150,170) -- (120,170) -- (120,110) -- (147,110) ;
\draw [shift={(150,110)}, rotate = 180] [fill={rgb, 255:red, 0; green, 0; blue, 0 }  ][line width=0.08]  [draw opacity=0] (8.93,-4.29) -- (0,0) -- (8.93,4.29) -- cycle    ;
\draw    (200,100) -- (247,100) ;
\draw [shift={(250,100)}, rotate = 180] [fill={rgb, 255:red, 0; green, 0; blue, 0 }  ][line width=0.08]  [draw opacity=0] (8.93,-4.29) -- (0,0) -- (8.93,4.29) -- cycle    ;
\draw    (320,110) -- (329.25,110) -- (367,110) ;
\draw [shift={(370,110)}, rotate = 180] [fill={rgb, 255:red, 0; green, 0; blue, 0 }  ][line width=0.08]  [draw opacity=0] (8.93,-4.29) -- (0,0) -- (8.93,4.29) -- cycle    ;
\draw    (430,100) -- (477,100) ;
\draw [shift={(480,100)}, rotate = 180] [fill={rgb, 255:red, 0; green, 0; blue, 0 }  ][line width=0.08]  [draw opacity=0] (8.93,-4.29) -- (0,0) -- (8.93,4.29) -- cycle    ;
\draw    (450,100) -- (450,170) -- (433,170) ;
\draw [shift={(430,170)}, rotate = 360] [fill={rgb, 255:red, 0; green, 0; blue, 0 }  ][line width=0.08]  [draw opacity=0] (8.93,-4.29) -- (0,0) -- (8.93,4.29) -- cycle    ;
\draw    (350,80) -- (367,80) ;
\draw [shift={(370,80)}, rotate = 180] [fill={rgb, 255:red, 0; green, 0; blue, 0 }  ][line width=0.08]  [draw opacity=0] (8.93,-4.29) -- (0,0) -- (8.93,4.29) -- cycle    ;
\draw   (320,150) -- (430,150) -- (430,190) -- (320,190) -- cycle ;
\draw    (320,170) -- (290,170) -- (263,170) ;
\draw [shift={(260,170)}, rotate = 360] [fill={rgb, 255:red, 0; green, 0; blue, 0 }  ][line width=0.08]  [draw opacity=0] (8.93,-4.29) -- (0,0) -- (8.93,4.29) -- cycle    ;

\draw (175,100) node   [align=left] {PID};
\draw (285,100) node   [align=left] {\begin{minipage}[lt]{46.95pt}\setlength\topsep{0pt}
\begin{center}
Thrust\\Allocation
\end{center}

\end{minipage}};
\draw (400,100) node   [align=left] {\begin{minipage}[lt]{32.78pt}\setlength\topsep{0pt}
\begin{center}
Vessel
\end{center}

\end{minipage}};
\draw (205,171) node   [align=left] {\begin{minipage}[lt]{62.81pt}\setlength\topsep{0pt}
\begin{center}
Extended \\Kalman Filter
\end{center}

\end{minipage}};
\draw (124.88,79.2) node  [font=\large] [align=left] {$\displaystyle \nu _{ref}$};
\draw (137,156) node  [font=\large] [align=left] {$\displaystyle \hat{\nu }$};
\draw (217.19,89.2) node  [font=\large] [align=left] {$\displaystyle \tau _{cmd}$};
\draw (344,97) node  [font=\large] [align=left] {$\displaystyle F_{i} ,\alpha _{i}$};
\draw (445.55,89) node  [font=\large] [align=left] {$\displaystyle \eta ,\nu $};
\draw (353.5,67) node  [font=\large] [align=left] {$\displaystyle b$};
\draw (375,170) node   [align=left] {\begin{minipage}[lt]{70.18pt}\setlength\topsep{0pt}
\begin{center}
Measurements
\end{center}

\end{minipage}};

\end{tikzpicture}